# Application of Computational Physics: Blood Vessel Constrictions and Medical Infuses

Suprijadi[*], M. R. A. Sentosa, P. Subekti, and S. Viridi

*Department of Physics, Faculty of Mathematics and Natural Sciences, Institut Teknologi Bandung, Jl. Ganesha No. 10 Bandung, Indonesia*
*Email: supri@fi.itb.ac.id

**Abstract.** Application of computation in many fields are growing fast in last two decades. Increasing on computation performance helps researchers to understand natural phenomena in many fields of science and technology including in life sciences. Computational fluid dynamic is one of numerical methods which is very popular used to describe those phenomena. In this paper we propose moving particle semi-implicit (MPS) and molecular dynamics (MD) to describe different phenomena in blood vessel. The effect of increasing the blood pressure on vessel wall will be calculate using MD methods, while the two fluid blending dynamics will be discussed using MPS. Result from the first phenomenon shows that around 80% of constriction on blood vessel make blood vessel increase and will start to leak on vessel wall, while from the second phenomenon the result shows the visualization of two fluids mixture (drugs and blood) influenced by ratio of drugs debit to blood debit.

## INTRODUCTION

Blood flow in medium and large vessels (heart, arteries, veins) can be considered as Newtonian fluid and neglects the influence of blood microstructure [1]. It means the Navier–Stokes equations are considered to be a good model for blood flow, although the blood is a suspension of red blood cells, white blood cells, and platelets in plasma, its non-Newtonian nature due to the particular rheology is only relevant in small arteries (arterioles) and capillaries where the diameter of the arteries becomes comparable to the size of the cells [2]. Also, a blood flow is governed by the continuity equation [3]. Motion of blood in blood vessels is one of major applications of natural motion. Understanding blood flow along arterial walls is important in understanding the mechanisms leading to various complications in the cardiovascular function.

In order to study blood flow, one of the available tools is fluid dynamics. Computational fluid dynamic is one of numerical methods which is very popular used to study fluid flow. Different methods were applied in recent years. Discrete model of simulation were studying, such as smoothed particle hydro-dynamic (SPH) [4] and MPS [5]. The application of these discrete models on natural motion are growing fast.

Many clinical treatments can be studied in detail only if a reliable model describing the response of arterial walls to the blood flow is considered [2]. On the other hand, clinical treatment such as infuses is interesting to study, the mixed two different fluid in a micro diameter tube such us blood vessel is not understand well, because of lack instrumentation to observed or detect the phenomena in living system.

In this paper we propose computation study on human blood vessel, at first section we study to understand the effect of constriction wall of blood vessel in bleeding phenomena, but a litte bit different than [6], and in second section study to understand the effect of velocity and infusion tube diameter using two fluids mixture dynamic computation.

## MOLECULAR DYNAMIC ON CONSTRICTED BLOOD VESSEL

In Lagrangian frame, the Navier-Stokes equations take form of Equations (1) and (2)

$$\frac{D\rho}{Dt} = -\rho \nabla \cdot \mathbf{v} , \qquad (1)$$

$$\frac{D\mathbf{v}}{Dt} = -\frac{\nabla p}{\rho} + \cdot \mathbf{g} + \mathbf{\Theta} , \qquad (2)$$

where $\rho$, $p$, $\mathbf{v}$ and $\mathbf{g}$ are density, pressure, velocity and body force (e.g. gravity), while $\Theta$ is refers to viscous term. As the Navier-Stokes equation was derived, the blood pressure at blood vessels are determined. Blood pressure is defined as the pressure experienced upon the walls of blood vessels during the circulation of blood, and this is one of the vital signs of physical health.

Now, in developing the model of blood pressure, Poiseuille's equation has been used, which determines the relation between blood flow rate and blood pressure, and it is given by Equation (3)

$$\Delta p = \frac{8\mu l Q}{\pi r^4}, \quad (3)$$

where, $l$ is the length of straight vessel and $r$ is the radius of blood vessels. $\mu$ is the kinematic viscosity of blood, $Q$ is the rate of blood flow, and $\Delta p$ is pressure difference.

## Artery wall model

The artery wall was formed by bond-particle of two atoms. Each near two atoms were connected by bent spring. The connection that formed each two atoms have a force, in this case is spring force, and specific constant. As we can see in Figure 1. Potential energy was kept in a bent spring, and write as equation in

$$E_p = \frac{1}{2}kx^2, \quad (4)$$

where $k$ is spring constant and $x$ is displacement of wall particles.

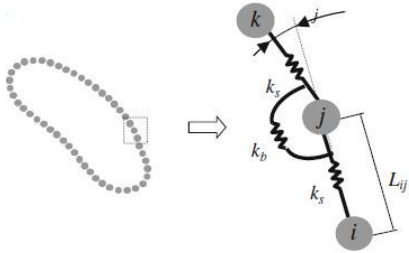

**FIGURE 1.** Artery wall formed by bond-particle that connected by bent spring [7].

In Equation (4), potential energy is a step increment energy that occurred because of spring compression. Artery wall was more complex and contained of two types of elastic energy, that is, elastic energy in stretch spring and elastic energy in bending spring [8]. Elastic energy that kept in stretch spring was defined as follows in Equation (5)

$$E_s = \frac{1}{2}k_s \sum_{j=1}^{N}\left(\frac{L_{ij}-L_0}{L_0}\right)^2, \quad (5)$$

where $L_{ij}$ is a element of two particles $i$ and $j$, $L_0$ is refence length, and $k_s$ is spring constant (N/m). For bending spring, elastic energy was defined as function of angle, $\theta_{ijk}$, between two line elements (or three particles) as in Equation (6)

$$E_b = \frac{1}{2}k_b \sum_{j=1}^{N}\tan^2\left(\frac{\theta_{ijk}}{2}\right)^2, \quad (6)$$

with $k_b$ is spring constant in bending spring (N/m).

Lennard-Jones potential is chosen for interactions between blood particles and between blood and wall particles) [8], which is given by Equation (7)

$$V_{LJ} = 4\varepsilon\left[\left(\frac{\sigma}{r}\right)^{12} - \left(\frac{\sigma}{r}\right)^{6}\right], \quad (7)$$

where $V_{LJ}$ is Lennard-Jones potential, $\varepsilon$ is well-depth and a measure of how strongly the two particles attract each other, $\sigma$ is the distance at which the intermolecular potential between the two particles is zero. It gives a measurement of how close two nonbonding particles can get. It is equal to one-half of the internuclear distance between nonbonding particles. And lastly, $r$ is the distance of separation between both particles (measured from the center of one particle to the center of the other particle). We use these kind of atomic potential for interactions between blood particles and between blood and vessel wall particles by considering why it can not, while it holds for other fluids [9].

## Pressure time series and leakage

The model was calculated using Molecular Workbench (a free MD software), with technical details can found in [10]. At initial condition, we put some gauge to measure the pressure at two different position, as can be seen in Figure 2 (top) denoted as A in inlet of constriction vessel, and B at outlet of vessel. The blood particle were represented by dots. At beginning, the blood flow from left to right direction with similar flow. After 60,000 fs some blood particle were leaking (an internal bleeding phenomenon) for ratio of constriction $c > 80$ %, as in Figure 2 (bottom), denoted by X. Number of particles depends on initial pressure given at inlet and $c$. In some area just after constriction vessel we observed some vacuum spot which is denoted by V.

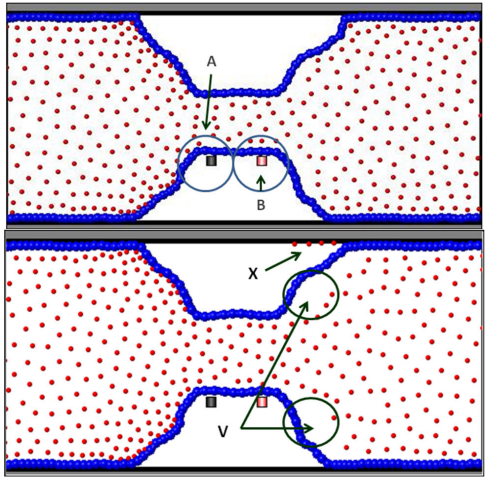

**FIGURE 2.** The simulation of blood flow in blood vessels: initial condition (top) and after 60.000 fs (bottom) for value of $c > 80$ %.

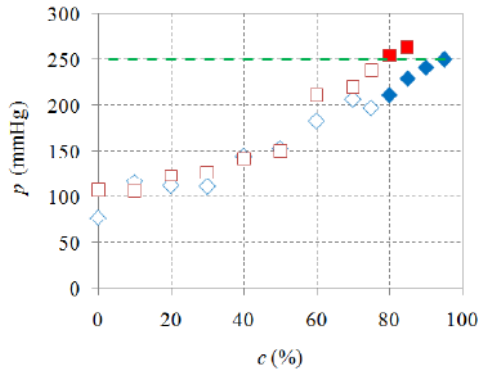

**FIGURE 3.** Measure of pressure p in constricted part of vessel as function of constriction ratio at different initial pressure of 105 mmHg (□) and 75 mmHg (◇) the solid markers mean that the blood particle leakage in some area like denote as X in Figure 2 (bottom).

Figure 3 shows the result of simulation in different constriction ratio from 10% up to 95 %. The different inlet pressure were applied with spring constant $k = 0.2$. Some leakage of blood particle as denoted by X in Fig.2 (bottom) were occur, specially at 80% of constriction. The leakage meaning that the pressure measured in constricted area more than 250 mmHg, this pressure threshold was confirmed by [11].

## MOVING PARTICLE SEMI IMPLICIT (MPS) ON MEDICAL INFUSES

The governing equations, Equations (1) and (2) can be separated into two different forces, it means external force which define by $\left(\frac{\partial u}{\partial t}\right)_{explicit} = a_{ext}$ and internal forces which define by $\left(\frac{\partial u}{\partial t}\right)_{implicit} = -\frac{1}{\rho}\nabla p$.

In this simulation, MPS will concern to analyze internal forces effect on its acceleration and each particle close to an particle subject will give effect to acceleration, velocity, and density of region, as well give different viscosity every step calculation [12]. Figure 4 shows the effect of surrounding particle into a particle subject (denote by $i$).

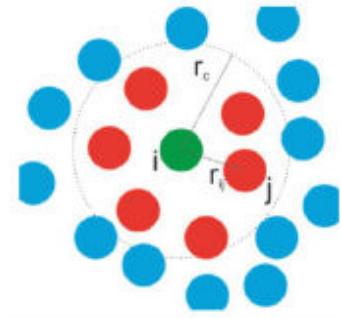

**FIGURE 4.** Particle $i$ is surrounding by many particle and threshold distance $r_c$.

The weight function of particle $i$ can be written as Equation (8)

$$w(r_{ij}) = \begin{cases} \frac{r_c}{r_{ij}} - 1, & 0 < r_{ij} \leq r_c, \\ 0, & r_{ij} > r_c. \end{cases} \quad (8)$$

### Blood - Drugs infuses model

In modeling of medical infuses, the system can be simplify as in Figure 5, the drugs particle parameter is defined in domain 2, which is can be controlled. The controlled parameters are viscosity, flow rate, and tube diameter. All of these parameters will be compared to blood particle parameters in domain 1. To simplify the simulation, all vessel walls are assumed by fixed particles and rigid. This assumption is very different than in the previous section simulation.

### Penetration of drugs flow

Figure 5 shows the simulation results from different ratio of drugs flow rate to the blood flow rate $R$. Increasing of ratio shows the drugs more penetrating into the flow in blood vessel. But in some how we did not find the drugs replace whole section of blood (good mixture) as founded in previous work [5]. The possibility of this condition is, different model was used, specially in vessel wall and the inlet model of blood. In our cases the inlet were define with continue flow and assume that the blood particle have the same pressure, flow, and viscosity in every step in domain 1.

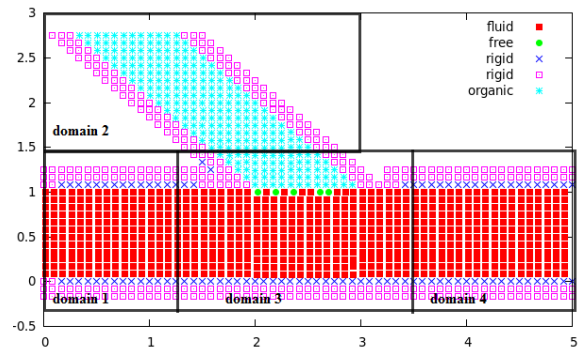

**FIGURE 5.** The two fluid model as represented to medical infuses, the blood vessel divided into 3 domain, denoted by domain 1, 3 and 4, while drug particle denoted by domain 2.

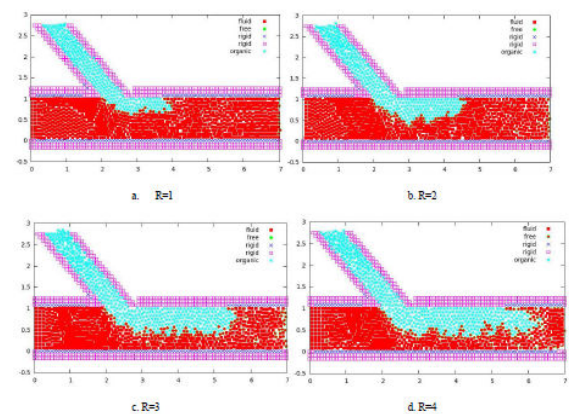

**FIGURE 6.** Simulation results as function of ratio of drugs debit to the blood debit $R$: (a) 1, (b) 2, (c) 3, and (d) 4.

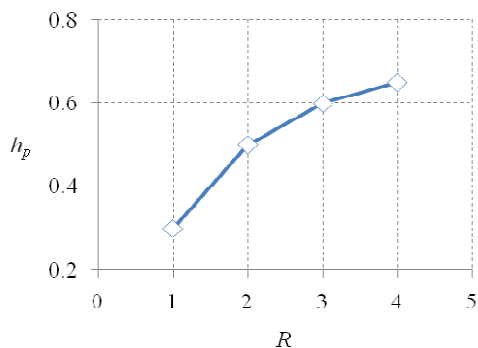

**FIGURE 7.** Penetration depth $h_p$ as function of ratio of drugs flow rate to blood flow rate $R$.

From the visualizations in Figure 6, penetration depth $h_p$, the distance where drugs flow can penetrate blood flow, as function of $R$ is obtained. The result is shown in Figure 7, which is reasonable comparing to common reported results for liquid holdup which depends on input liquid content [13].

## CONCLUSION

Application of molecular dynamic (MD) and moving particle semi implicit (MPS) to blood simulation were reported: (i) the effect of blood vessel constriction can be understood, that the blood leakages occur when constriction is about and more than 80 % in blood vessel; (ii) the drugs infuses flow rate plays important role on penetration of drugs flow into blood flow, as it shown in the simulation.